\documentclass[structabstract]{aa}

\usepackage{natbib}
\bibpunct{(}{)}{;}{a}{}{,}
\usepackage{multirow}
\usepackage{multicol}
\usepackage{slashbox}
\usepackage{amsmath}
\usepackage{amssymb}
\usepackage{wasysym}
\usepackage{epsfig}
\usepackage{epsf}
\usepackage{units}
\usepackage[usenames,dvipsnames,table]{xcolor}
\usepackage{graphicx,graphics}
\usepackage{float}
\usepackage{setspace,lineno}

\usepackage{txfonts}

\begin{document}
\begin{multicols}{1}

\title{Atmospheric studies of habitability in the Gliese 581 system}
\titlerunning{Habitability studies of GL 581 planets}

\author{P. von Paris\inst{1} \and S. Gebauer\inst{2} \and M. Godolt\inst{2} \and H. Rauer\inst{1,2} \and B. Stracke\inst{1}}

\institute{Institut f\"{u}r Planetenforschung, Deutsches Zentrum
f\"{u}r Luft- und Raumfahrt, Rutherfordstr. 2, 12489 Berlin, Germany
\and Zentrum f\"{u}r Astronomie und Astrophysik, Technische
Universit\"{a}t Berlin, Hardenbergstr. 36, 10623 Berlin, Germany}

\abstract {The M-type star \object{Gliese 581} is orbited by at
least one terrestrial planet candidate in the habitable zone, i.e.
GL 581 d. Orbital simulations have shown that additional planets
inside the habitable zone of GL 581 would be dynamically stable.
Recently, two further planet candidates have been claimed,
one of them in the habitable zone.}{In view of the ongoing
search for planets around M stars which is expected to result in
numerous detections of potentially habitable Super-Earths, we take
the GL 581 system as an example to investigate such
planets. In contrast to previous studies of habitability in
the GL 581 system, we use a consistent atmospheric model to assess surface conditions and habitability. Furthermore, we perform detailed atmospheric simulations for a much larger subset of potential planetary
and atmospheric scenarios than previously considered. } {A 1D radiative-convective
atmosphere model is used to calculate temperature and
pressure profiles of model atmospheres, which we assumed to be
composed of molecular nitrogen, water, and carbon dioxide. In these
calculations, key parameters such as surface pressure and CO$_2$
concentration as well as orbital distance and planetary
mass are varied.}
 { Results imply that surface temperatures above freezing could be obtained, independent of the here considered atmospheric scenarios,
at an orbital distance of 0.117 AU. For an orbital distance of 0.146
AU, CO$_2$ concentrations as low as 10 times the present Earth's
value are sufficient to warm the surface above the freezing point of
water. At 0.175 AU, only scenarios with CO$_2$ concentrations of 5
\% and 95 \% were found to be habitable. Hence, an additional
Super-Earth planet in the GL 581 system in the previously determined
dynamical stability range would be considered a potentially
habitable planet. }{}

\keywords{Astrobiology, Planets and satellites: atmospheres, Stars:
planetary systems, Stars: individual: \object{Gliese 581}}

\maketitle

\end{multicols}{1}

\section{Introduction}

The discovery and characterization of extrasolar terrestrial planets
in the habitable zone (HZ) of their central star is one of the most
exciting prospects of exoplanetary science. Such planets are
extremely good candidates for the search for extraterrestrial life.
The HZ is usually defined as the shell around a star where a planet
could retain liquid water on the surface \citep{kasting1993}. This
definition is motivated because liquid water seems to be the
fundamental requirement for life as we know it on Earth. Being
located inside the HZ as defined by \citet{kasting1993} for an
Earth-like planet, however, not necessarily implies habitability for
a specific planetary scenario (see, e.g., Mars in our own solar
system). The potential habitability of a planet depends critically
on atmospheric composition and surface pressure. Still, for a planet
located well inside this classical HZ, habitability is achievable
for a much broader range of atmospheric conditions (greenhouse
effect etc.) than for a planet near one of the boundaries.

Among the more than 500 extrasolar planets discovered so far, some
orbit their central star inside or near the HZ (e.g.,
\citealp{mayor2004}, \citealp{lovis2006}, \citealp{fischer2008},
\citealp{haghighipour2010}). Most of these planets are Neptune- or
Jupiter-like gas planets. The planetary system \object{Gliese 581}
(GL 581), however, contains at least four planets
(\citealp{bonfils2005}, \citealp{udry2007},
\citealp{mayor2009gliese}), one of which is a potentially habitable
Super-Earth, GL 581 d. This was shown by \citet{wordsworth2010},
\citet{vparis2010gliese}, \citet{hu2011} and
\citet{kaltenegger2011} who presented 1D modeling studies of
different atmospheric scenarios of GL 581 d. They found habitable
surface conditions (i.e., surface temperatures above 273 K) with
CO$_2$ partial pressures as low as 1 bar, depending on CO$_2$
concentration. These results imply that the GL 581 planetary system
contains indeed at least one potentially habitable, possibly
terrestrial planet.

Orbital simulations presented by \citet{zollinger2009}
showed that between the orbits of GL 581 c and d (i.e., inside the
classical HZ), another Super-Earth planet would be dynamically
stable. Specifically, they stated a stability range for a
low-eccentricity planet of not more than 2.6 Earth masses
(m$_{\oplus}$) ranging from 0.126 AU to 0.17 AU.

Recently, \citet{vogt2010gliese} claimed the detection of two more
planets in the GL 581 system, one of them (called GL 581 g in \citealp{vogt2010gliese}) with a minimum mass of 3.1
m$_{\oplus}$ and an orbital distance of 0.146 AU, hence
inside the stability range calculated by
\citet{zollinger2009}. These detections are controversial and
disputed by further analysis of the radial velocity data
\citep{tuomi2011}.

Nevertheless, we use this claimed discovery as a starting
point to investigate the habitability of planets in the GL 581
system. Such model calculations aim at supporting the selection of
future targets for detailed observational programs of habitable
planets which is probably needed in the future \citep{horner2010},
given the expected number of targets. Such potentially habitable
planets are expected to be discovered in the near future by on-going
ground-based programs such as MEarth \citep{nutzman2008} or space
missions like Kepler (see, e.g.,
\citealp{borucki2011} for an overview of Kepler candidates) and the
planned PlaTO mission \citep{catala2009}. First attempts at
characterizing the atmospheres of transiting Super-Earth planets
have already been made (CoRoT-7 b, \citealp{guenther2011}, and GJ
1214 b, \citealp{bean2010}, \citealp{desert2011}, \citealp{croll2011_gj1214}).

For the claimed planet in the HZ of GL 581, dedicated
modeling studies have been performed by \citet{pierrehumbert2011},
 \citet{heng2011} and \citet{bloh2011}. \citet{pierrehumbert2011} presented
several possible atmospheric scenarios (airless planet, pure N$_2$,
mixed CO$_2$/H$_2$O atmospheres) and discussed potential
implications for surface conditions, without detailed calculations
of the atmospheric structure for the mixed CO$_2$/H$_2$O cases. On the other hand,
\citet{heng2011} used a general circulation model of Earth to
simulate the dynamics and circulation on GL 581 g, however did not
investigate surface conditions and habitability in detail.
The study of \citet{bloh2011} used a geodynamic box model to
assess planetary habitability, coupling geophysical and atmospheric
processes in a simplified approach.

Previous modeling studies of habitability in the GL 581 system either focused on the existing planets GL 581 c and d (e.g., \citealp{selsis2007gliese}, \citealp{bloh2007}), used a very simple model to simulate atmospheric processes and surface conditions (e.g., \citealp{bloh2011}) or investigated only a very small subset of potential atmospheric scenarios in terms of CO$_2$ level and surface pressure when varying orbital distance \citep{kaltenegger2011}. We present here model calculations for
possible terrestrial planets in the GL 581 system
along the same line of reasoning as in \citet{vparis2010gliese}, using a consistent 1D atmosphere model. We vary the surface pressure and CO$_2$
level over a large range. For these scenarios, we calculate temperature and pressure profiles in order to
assess the habitability of up to now hypothetical
planets, assuming different planetary masses and orbital distances.

The paper is organized as follows: Sect. \ref{plansys} states the
stellar and planetary parameters. The model used is described in
Sect. \ref{model}. A description of the runs is given in Sect.
\ref{modinput}. Results are described and discussed in Sect.
\ref{resultsect}. We give our conclusions in Sect. \ref{concl}.


\section{Model planets around GL 581}
\label{plansys}

GL 581 is a very quiet M3 star \citep{bonfils2005}. The stellar
spectrum of GL 581 is taken from \citet{vparis2010gliese}. It was
derived from an UV spectrum measured by the IUE (International
Ultraviolet Explorer) satellite and a synthetic Nextgen model
spectrum \citep{hauschildt1999}.

Atmospheric simulations were performed for a subset of
probable planet scenarios, defined by mass and orbital distance. We
varied the orbital distance from 0.117 to 0.175 AU, covering the
stability range found by \citet{zollinger2009}. In terms of
insolation in the solar system, this translates roughly into the
present-day insolation at the orbit of Mars and the solar flux at
Earth about 1.3 billion years ago (e.g., \citealp{gough1981}). The
planetary mass was varied between 2.6 and 3.1 m$_{\oplus}$. For all
scenarios, orbits were assumed to be circular.

The planetary radius is taken from a mass-radius relationship by
\citet{sotin2007}, yielding the surface gravity. Changing
the planetary mass from 3.1 to 2.6 m$_{\oplus}$ decreases the
gravity from 16.4 ms$^{-2}$ to 15.1 ms$^{-2}$, i.e by roughly 9 \%.

As in \citet{vparis2010gliese}, the measured Earth surface albedo,
i.e. the reflectivity of the planetary surface with respect to
incoming stellar radiation, was taken as the model surface albedo
($A_{\rm{surf}}$=0.13, \citealp{rossow1999}) for all
scenarios. In doing so, the effect of clouds is explicitly excluded
from our simulations. Table \ref{planpar} summarizes the planetary
parameters. Note that scenario 2 in Table \ref{planpar}
corresponds to the claimed planet GL 581 g of \citet{vogt2010gliese}.

\begin{table}[H]

\caption{Planetary parameters}

\label{planpar}

\begin{center}
\resizebox{\hsize}{!}{\begin{tabular}{lrr}

  \hline
                      &  Mass [m$_{\oplus}$]    & Orbital distance [AU]             \\
  \hline
  Scenario 1          &  3.1                    & 0.117 \\
  Scenario 2          &  3.1                    & 0.146\\
  Scenario 3          &  3.1                    & 0.175\\
  Scenario 4          &  2.6                    & 0.146\\
\end{tabular}}
\end{center}

\end{table}

\section{Atmospheric model}

\label{model}

A cloud-free 1D radiative-convective model was used to calculate the
atmospheric structure, i.e. the temperature, water and pressure
profiles.

The model is originally based on the climate model described by
\citet{kasting1984water} and \citet{kasting1984}. Further
developments are described by e.g. \citet{kasting1988} and
\citet{mischna2000}.  The model version used here is based on the
version of \citet{vparis2008} and \citet{vparis2010gliese} where
more details on the model are given.

The model considers N$_2$, H$_2$O, and CO$_2$ as atmospheric
species. H$_2$O and CO$_2$ are the two most important greenhouse
gases on present Earth, and N$_2$ is present in significant amounts
in all terrestrial atmospheres of the solar system.

Temperature profiles from the surface up to a pressure of 6.6
$\cdot$ 10$^{-5}$ bar are calculated by solving the equation of
radiative transfer and performing convective adjustment, if
necessary. Convective adjustment means that the lapse rate in the
atmosphere is adjusted to the convective lapse rate instead of using
the radiative lapse rate if the atmosphere is unstable against
convection. The convective lapse rate is assumed to be adiabatic
with contributions of latent heat release by condensing water or
carbon dioxide. The water profile is calculated based on the
relative humidity distribution of \citet{manabewetherald1967}. Above
the cold trap, the water profile is set to an isoprofile of the cold
trap value.

\section{Atmospheric scenarios}

\label{modinput}

We performed a parameter study to investigate the influence of
surface pressure and CO$_2$ level on the potential habitability of
terrestrial planets in the GL 581 system, as summarized in
Table \ref{planpar}. The initial surface pressure (1, 2, 5, 10, 20
bar) and CO$_2$ concentration (0.95, 0.05, 3.55 $\cdot$ 10$^{-3}$
and 355 ppm, respectively) were varied. N$_2$ was used as the
background gas. Table \ref{listofruns} summarizes the considered
atmospheric scenarios.

\begin{table}[H]
  \caption{Atmospheric scenarios (PAL: Present
  Atmospheric Level)
  }\label{listofruns}
\begin{center}
\resizebox{\hsize}{!}{\begin{tabular}{lcc}
 \hline
   Set  &   $p$ [bar]     & CO$_2$ vmr           \\
  \hline
    G1 (low CO$_2$)  & 1,2,5,10,20       &3.55 $\cdot$ 10$^{-4}$           \\
    G2 (10 PAL CO$_2$)  & 1,2,5,10,20       &3.55 $\cdot$ 10$^{-3}$           \\
    G3 (medium CO$_2$) & 1,2,5,10,20       &0.05                              \\
    G4  (high CO$_2$)& 1,2,5,10,20       &0.95                       \\
\end{tabular}}
\end{center}
\end{table}

\section{Results and discussion}

\label{resultsect}

\subsection{Temperature profiles}

\begin{figure}[H]
\includegraphics[width=200pt]{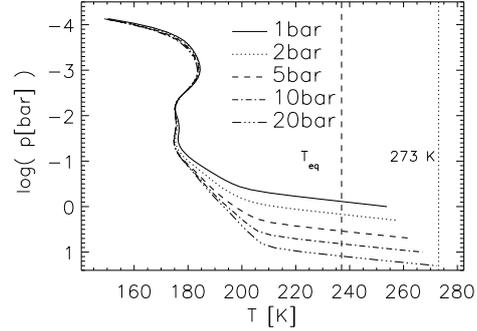}\\
  \caption[Temperature-pressure profiles for the low CO$_2$ runs]
  {Temperature-pressure profiles for the low CO$_2$ (355 ppm CO$_2$) runs of scenario 2.
  Equilibrium temperature of the planet (dashed) and melting temperature of water (dotted) are indicated as vertical lines. }
  \label{temperature_low}
\end{figure}

\begin{figure}[H]
\includegraphics[width=200pt]{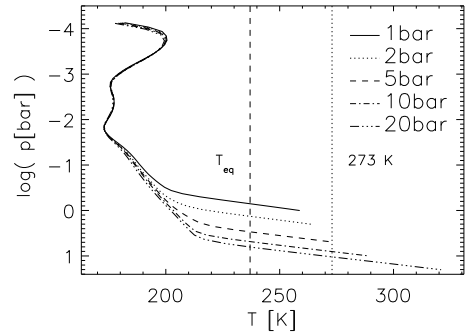}\\
  \caption[Temperature-pressure profiles for the 10 PAL CO$_2$ runs]
  {Temperature-pressure profiles for the 10 PAL CO$_2$ (3.55 $\cdot$ 10$^{-3}$ CO$_2$) runs of scenario 2.
  }
  \label{temperature_pal}
\end{figure}

\begin{figure}[H]
\includegraphics[width=200pt]{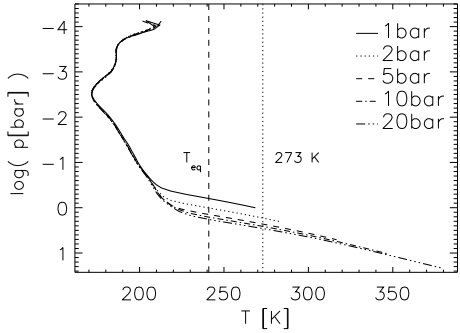}\\
  \caption[Temperature-pressure profiles for the medium CO$_2$ runs]
  {Temperature-pressure profiles for the medium CO$_2$ (5\% CO$_2$) runs of scenario 2.
  }
  \label{temperature_medium}
\end{figure}

\begin{figure}[H]
 \includegraphics[width=200pt]{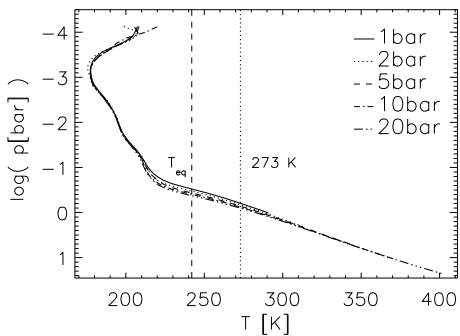}\\
  \caption[Temperature-pressure profiles for the high CO$_2$ runs]
  {Temperature-pressure profiles for the high CO$_2$ (95\% CO$_2$) runs of scenario 2.}
  \label{temperature_high}
\end{figure}

The resulting temperature-pressure profiles of scenario 2 in
Table \ref{planpar} for the sets G1-G4 are shown in Figs.
\ref{temperature_low}-\ref{temperature_high}. The equilibrium
temperature $T_{\rm{eq}}$ of the planet and the melting temperature
of water (273 K) are indicated by vertical lines. A global mean
surface temperature of 273 K or higher is generally used as the
criterion for surface habitability in exoplanet science. This
criterion is purely based on the phase diagram of water where the
liquid phase of water needs temperatures above 273 K at almost all
pressures (the melting line is nearly isothermal in the  p-T
diagram). Note that, of course, on Earth life is found in areas with
mean annual temperatures far below the freezing point of water.

The low CO$_2$ runs result in uninhabitable surface conditions for
all assumed surface pressures (see Fig. \ref{temperature_low}),
although the 20 bar run with a surface temperature of 272 K is very
close to being considered habitable.

When increasing the CO$_2$ content by a factor of 10, calculated
surface temperatures were ranging between 258 and 320 K (Fig.
\ref{temperature_pal}). For surface pressures of 5 bar or more,
surface temperatures were larger than 273 K, indicating potentially
habitable surface conditions. The rather high surface temperature of
320 K for the 20 bar run is the result of the stronger greenhouse
effect due to CO$_2$ and a positive water vapor feedback (increasing
surface temperature leads to increased water vapor in the
atmosphere, hence more greenhouse effect).

The medium CO$_2$ runs (see Fig. \ref{temperature_medium}) result in
habitable conditions on the surface for pressures of 2 bar and
higher. Calculated surface temperatures are as high as 378 K for the
20 bar case.

The high CO$_2$ runs (see Fig. \ref{temperature_high}) all show
habitable conditions on the surface. Calculated surface temperatures
range from 290 to 401 K upon increasing the surface pressure from 1
to 20 bar.

In all runs presented here, model atmospheres show convective
tropospheres and radiative stratospheres, in contrast to purely
radiative atmospheres encountered in some cases in
\citet{vparis2010gliese}. Also, CO$_2$ condensation is absent for
all runs, even in the high CO$_2$ cases, due to the high
temperatures in the middle atmosphere.

\subsection{Variation of gravity}

The main effect upon decreasing the gravity $g$ at fixed
surface pressure $p_s$ is an increase in column density $D_{\rm{col}}$ (since 
$D_{\rm{col}}\sim \frac{p_s}{g}$). This then translates into more greenhouse
effect, hence the first-order influence of decreasing gravity will
be an increase in surface temperature.\newline However, for the
small variations in gravity considered here (see Sect.
\ref{plansys}), the resulting increase in surface temperatures was
only of the order of 0.4 to 2.8 K. Such small variations of surface
temperatures are not critical in the assessment of habitability in
the frame of this work and are therefore not shown.

\subsection{Variation of orbital distance}

The immediate effect of increasing the orbital distance is a
reduction of the stellar flux $S$ received by the planet. Hence, the
amount of CO$_2$ required to achieve surface temperatures above 273
K increases with orbital distance.

In Table \ref{table_summary_results}, we show the surface temperatures as a
function of surface pressure and CO$_2$ concentration for the scenarios 1-3 
in Table \ref{planpar}.\newline It is clearly seen that for an orbital distance of 0.117 AU even the low CO$_2$ runs
result in habitable surface conditions over the entire range of
surface pressures considered. It can also be inferred that
the same value of surface temperature can be achieved for several
combinations of surface pressure and CO$_2$ concentration.\newline
As was already demonstrated by the temperature
profiles shown above, for an orbital distance of 0.146 AU, several model scenarios resulted in surface
conditions which were uninhabitable (e.g., Fig.
\ref{temperature_low}). However, in general, habitability can be
achieved over the entire range of CO$_2$ concentrations with accordingly high surface pressures.

\begin{table}[H]
\caption{Surface temperature in K for variation of orbital distance $d$, surface pressure $p$ in bar and CO$_2$ concentration $C$. Increased grey shading between 260 and 380 K in 30 K steps. White indicates cold scenarios.}\label{table_summary_results}
\begin{tabular}{cp{0.7cm}p{0.7cm}p{0.7cm}p{0.7cm}p{0.7cm}}
\hline\hline
\backslashbox{$C$}{$p$} & 1        & 2&5        & 10      & 20\\\hline
\multicolumn{6}{c}{$d$=\unit[0.117]{AU}}\\\hline
\unit[355]{ppm}     &\cellcolor[gray]{0.9}288 &\cellcolor[gray]{0.8}296  &\cellcolor[gray]{0.8}309   &\cellcolor[gray]{0.7}323    &\cellcolor[gray]{0.7}341\\
\unit[3550]{ppm}   & \cellcolor[gray]{0.8}295&\cellcolor[gray]{0.8}305 & \cellcolor[gray]{0.7}326 &\cellcolor[gray]{0.7}347 &\cellcolor[gray]{0.6}372\\
\unit[5]{\%}       &  \cellcolor[gray]{0.8}307 &\cellcolor[gray]{0.7}324   &\cellcolor[gray]{0.6}351&\cellcolor[gray]{0.6}375 &\cellcolor[gray]{0.5}404\\
\unit[95]{\%}      & \cellcolor[gray]{0.8}320&\cellcolor[gray]{0.7}340&\cellcolor[gray]{0.6}369&\cellcolor[gray]{0.5}395&\cellcolor[gray]{0.5}424\\

\hline
\multicolumn{6}{c}{$d$=\unit[0.146]{AU}}\\
\hline
\unit[355]{ppm}     &254 & 257  &\cellcolor[gray]{0.9}263   &\cellcolor[gray]{0.9}267    &\cellcolor[gray]{0.9}272\\
\unit[3550]{ppm}   & 259& \cellcolor[gray]{0.9}264&\cellcolor[gray]{0.9} 274 &\cellcolor[gray]{0.9}289 &\cellcolor[gray]{0.7}321\\
\unit[5]{\%}       &  \cellcolor[gray]{0.9}269 & \cellcolor[gray]{0.9}283  &\cellcolor[gray]{0.8}317&\cellcolor[gray]{0.7}347 &\cellcolor[gray]{0.6}378\\
\unit[95]{\%}      & \cellcolor[gray]{0.8}291&\cellcolor[gray]{0.8}313&\cellcolor[gray]{0.7}345&\cellcolor[gray]{0.6}372&\cellcolor[gray]{0.5}401\\

\hline
\multicolumn{6}{c}{$d$=\unit[0.175]{AU}}\\\hline
\unit[355]{ppm}     & 221                &223         &225                &227                 &231\\
\unit[3550]{ppm}   & 224                &226         &232                &242                  &260\\
\unit[5]{\%}       &230                 &240         &\cellcolor[gray]{0.9}270&\cellcolor[gray]{0.8}309 & \cellcolor[gray]{0.7}347\\
\unit[95]{\%}      &254                 &\cellcolor[gray]{0.9}281&\cellcolor[gray]{0.8}320&\cellcolor[gray]{0.7}350 &\cellcolor[gray]{0.5}381\\
\hline
		\end{tabular}		
\end{table}

Also towards the outer boundary of
the HZ around GL 581, at an orbital distance of 0.175 AU, habitable scenarios could be
found. However, allowed CO$_2$ concentrations for surface temperatures above freezing are now limited to
the medium and high CO$_2$ case. Model scenarios with less CO$_2$
were found to be uninhabitable, independent of surface
pressure.\newline These results confirm the findings of
\citet{vparis2010gliese} that medium CO$_2$ scenarios need to be
taken into account when assessing the habitability of planets
orbiting near the outer boundary of the habitable zone.

\section{Conclusions}

We presented 1D radiative-convective calculations for a
subset of potential atmospheric conditions on hypothetical
Super-Earth planets orbiting GL 581. We varied parameters such as
orbital distance, CO$_2$ concentration and surface pressure. In contrast to previous studies of habitability 
in the GL 581 system, we considered a much larger parameter space and used a consistent atmospheric 
model to assess surface conditions.

Our model results imply that habitable surface
conditions (here T$_{\rm{surf}}>$273 K) could be obtained for a
large part of the considered parameter space. For the smallest orbital 
distance of 0.117 AU, habitability was achieved
independent of considered atmospheric parameter range. For an orbital distance 
of 0.146 AU, depending on surface pressure, CO$_2$
concentrations as low as 10 times the present Earth's value were
found to be sufficient for surface habitability. However, for the
largest orbital distance considered (0.175 AU), surface conditions were only habitable with
CO$_2$ concentrations of 5 \% and more. Hence, our simulations show
that an additional Super-Earth planet in the GL 581 system in the
dynamical stability range calculated by \citet{zollinger2009} would
indeed be considered a potentially habitable planet.

The model calculations presented here for a subset of the 
possible parameter space (orbital distance, CO$_2$ concentration,
surface pressure) illustrate how such investigations can be helpful
to select potentially habitable planets for further, more detailed
studies from a future larger sample of known Super-Earths.

\label{concl}

\begin{acknowledgements}

This research has been supported by the Helmholtz Gemeinschaft
through the research alliance "Planetary Evolution and Life".

Helpful discussions with J.W. Stock, J.L. Grenfell, A.B.C. Patzer and A.
H\"{o}lscher are gratefully acknowledged.

We thank the anonymous referee and Tristan Guillot for their
comments which helped clarify the paper.

\end{acknowledgements}

\bibliographystyle{aa}
\bibliography{literatur_phd}


\end{document}